\documentclass[conference]{IEEEtran}
\IEEEoverridecommandlockouts
\usepackage{cite}
\usepackage{amsmath,amssymb,amsfonts}
\usepackage{algorithmic}
\usepackage{graphicx}
\usepackage{textcomp}
\usepackage{xcolor}
\usepackage{times}
\usepackage{soul}
\usepackage{url}
\usepackage[hidelinks]{hyperref}
\usepackage[utf8]{inputenc}
\usepackage[small]{caption}
\usepackage{graphicx}
\usepackage{amsmath}
\usepackage{booktabs,bm}
\usepackage{algorithm}
\usepackage{algorithmic}
\usepackage{epsfig}
\usepackage{color}
\usepackage{dirtytalk}
\usepackage{subfigure}
\usepackage{multirow}
\usepackage{epstopdf}
\usepackage{setspace}
\usepackage{enumitem}
\usepackage{float}
\usepackage{wrapfig}
\usepackage{multirow}
\usepackage{capt-of}
\def\BibTeX{{\rm B\kern-.05em{\sc i\kern-.025em b}\kern-.08em
    T\kern-.1667em\lower.7ex\hbox{E}\kern-.125emX}}
\begin{document}

\title{RTMobile: Beyond Real-Time Mobile Acceleration of RNNs for Speech Recognition}

\author{\IEEEauthorblockN{Peiyan Dong,\IEEEauthorrefmark{1}
Siyue Wang,\IEEEauthorrefmark{1}
Wei Niu,\IEEEauthorrefmark{2}
Chengming Zhang,\IEEEauthorrefmark{3} 
Sheng Lin,\IEEEauthorrefmark{1}
Zhengang Li,\IEEEauthorrefmark{1}
Yifan Gong,\IEEEauthorrefmark{1} \\
Bin Ren,\IEEEauthorrefmark{2}
Xue Lin,\IEEEauthorrefmark{1}
Yanzhi Wang,\IEEEauthorrefmark{1}
and Dingwen Tao\IEEEauthorrefmark{3}\thanks{Corresponding author: Dingwen Tao, Department of Computer Science, The University of Alabama, Tuscaloosa, AL 35487, USA.}
}
\IEEEauthorblockA{\IEEEauthorrefmark{1}Northeastern University, MA, USA\\
\{dong.pe, wang.siy, lin.sheng, li.zhen\}@husky.neu.edu, \{xue.lin, yanz.wang\}@northeastern.edu}
\IEEEauthorblockA{\IEEEauthorrefmark{1}The College of William and Mary, VA, USA\\ wniu@email.wm.edu, bren@cs.wm.edu}
\IEEEauthorblockA{\IEEEauthorrefmark{3}
The University of Alabama, AL, USA\\ czhang82@crimson.ua.edu, tao@cs.ua.edu}
}


\maketitle

\begin{abstract}


Recurrent neural networks (RNNs) based automatic speech recognition has nowadays become prevalent on mobile devices such as smart phones. However, previous RNN compression techniques either suffer from hardware performance overhead due to irregularity or significant accuracy loss due to the preserved regularity for hardware friendliness. In this work, we propose \textit{RTMobile} that leverages both a novel block-based pruning approach and compiler optimizations to accelerate RNN inference on mobile devices. Our proposed RTMobile is the first work that can achieve real-time RNN inference on mobile platforms.
Experimental results demonstrate that RTMobile can significantly outperform existing RNN hardware acceleration methods in terms of inference accuracy and time. Compared with prior work on FPGA, RTMobile using Adreno 640 embedded GPU on GRU can improve the energy-efficiency by about 40$\times$ while maintaining the same inference time.

\end{abstract}

\begin{IEEEkeywords}
RNN, pruning, real-time acceleration, mobile
\end{IEEEkeywords}

\section{Introduction}
Deep neural network (DNN) has evolved to the state-of-the-art technique due to its high prediction accuracy in many artificial intelligence tasks, such as image recognition and characterization~\cite{krizhevsky2009learning,he2016deep,lecun2015lenet,krizhevsky2012imagenet,wang2018defensive,wang2019protecting,zhao2017aircraft}, speech recognition~\cite{graves2013speech,deng2013new,collobert2008unified,li2019rnn}, and recommender system \cite{wang2015collaborative}.
Among various DNN architectures, recurrent neural networks (RNNs) are widely used for speech recognition tasks because they can contain cycles to carry information across neurons when reading inputs. For instance, Gated Recurrent Unit (GRU)~\cite{cho2014properties}, the most recent representative popular type of RNNs, achieve great success in automatic speech recognition.
In recent years, executing DNNs on mobile platforms has become more and more popular because many high-end mobile devices are emerging. 
Several recent studies have proposed techniques to accelerate large-scale DNNs in mobile environment.  
However, due to fairly high computation complexity and memory consumption when executing RNNs, it is very challenging to deploy RNNs on current embedded processors and mobile devices to achieve real-time inference.

DNN model compression provides an effective way to mitigate the computation and memory challenges bringing by DNNs \cite{jin2019deepsz}. 
Many model compression techniques have been studied for recent years. For example, weight pruning can provide a notable reduction ratio in the model size. Early work~\cite{han2015deep} on non-structured weight pruning eliminates weights at arbitrary location, which leads to the pruned model to be stored in a sparse matrix format, such as compressed sparse column (CSC) format.
Non-structured weight pruning, however, hurts processing throughput because the indices in the compressed weight representation result in stalls or complex workloads on highly parallel architectures, such as GPUs and FPGAs.
On the other hand, structured weight pruning~\cite{wen2016learning} is more hardware friendly. By exploiting filter pruning~\cite{he2018soft} and channel pruning~\cite{he2017channel}, the pruned model is more regular in terms of the shape, which can eliminate storing the weight indices. However, structured pruning hurts accuracy more than non-structured pruning.
Moreover, state-of-the-art model-compression-based RNN acceleration techniques such as ESE~\cite{han2017ese} and C-LSTM~\cite{wang2018c} still suffer from limited inference accuracy and processing throughput, which prevents them to be implemented on mobile devices. 
Furthermore, existing DNN acceleration frameworks for mobile devices such as TVM~\cite{chen2018tvm} \textit{do not even support RNN}. 
Therefore, in order to achieve the real-time inference for RNNs on mobile devices, it is necessary to develop an end-to-end RNN acceleration framework that can achieve both high inference accuracy and high computational efficiency.

In this paper, we propose a real-time RNN acceleration framework for mobile devices named \textit{RTMobile}.
RTMobile is composed of two main components: block-based structured pruning and compiler-assisted performance optimization. 
Unlike traditional structured pruning methods used on DNNs, our novel block-based structured pruning approach that can provide a finer pruning granularity to maintain high inference accuracy while significantly reducing the RNN model size.
We also propose several compiler-based optimization techniques to determine the block size and generate the optimal code on mobiles. Our contributions are summarized as follows.
\begin{itemize}
    \item We propose a novel RNN acceleration framework for mobile devices, namely, RTMobile. To the best of our knowledge, \textbf{RTMobile is the first work that achieves real-time RNN inference on mobile devices}.
    \item We propose a fine-grained \underline{B}lock-based \underline{S}tructured \underline{P}runing algorithm (BSP) 
    for both high inference accuracy and high computational efficiency.
    \item We develop a series of compiler-based optimization techniques to further accelerate RNN inference on mobile platforms, including matrix reorder, load redundant elimination, and a new compact data format for pruned model storage (called BSPC, i.e., \underline{B}lock-based \underline{S}tructured \underline{P}runing \underline{C}ompact format). 
    \item We compare RTMobile with multiple state-of-the-art methods based on a representative RNN (GRU) using a well-known speech recognition dataset. Evaluation results demonstrate that RTMobile is \textbf{the first work that can compress the GRU model by over 10x without losing accuracy}. Experiments also illustrate that RTMobile can obtain about \textbf{50x energy-efficiency improvement} over prior work with the same inference time.
\end{itemize}


\section{Background and Motivation}
\label{sec:background}

In this section, we present some background information about GRU, DNN model compression, and DNN mobile acceleration framework, and discuss our research motivation.

\subsection{Gated Recurrent Unit}

\begin{figure}
\vspace{-5mm}
\begin{center}
	\includegraphics[width = 0.47\textwidth]{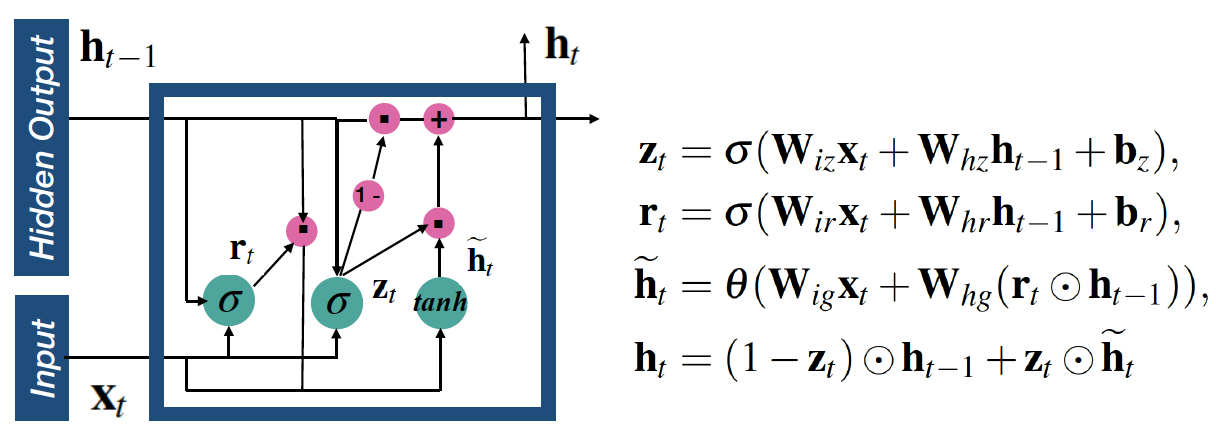}
	\caption{A Single GRU model.}
	\label{fig:GRU}
\end{center}
\vspace{-8mm}
\end{figure}

The \emph{\textbf{Gated Recurrent Unit (GRU)}} is a variation from the LSTM, proposed by Cho et al.\cite{cho2014properties}. 
It combines the forget and input gates into a single ``update gate''.
It also merges the cell state and hidden state, and makes some other changes.
The resulting GRU model is simpler than standard LSTM models, and has been growing increasingly popular.
Fig. \ref{fig:GRU} shows a single GRU, whose functionality is derived by using the following equations iteratively from $t = 1$ to $T$,
where symbols $\mathbf{z}$, $\mathbf{r}$, $\mathbf{\widetilde h}$, $\mathbf{h}$ are respectively the update gate, output gate, cell state, and cell output. As GRU is a more advanced version of RNN than LSTM, we mainly focus on GRU model in this work.

\subsection{DNN Model Compression Techniques}
As a representative technique in DNN model compression, DNN weight pruning removes the redundant or less important weights to reduce the storage and computational costs for the inference phase. There exist two mainstreams of weight pruning, i.e., non-structured pruning and structured pruning. 

\paragraph{Non-structured pruning}  Non-structured weight pruning is fine-grained and prunes weights at arbitrary locations. The early work proposed by Han et al. \cite{han2015learning} leverages a heuristic method to iteratively prune weights with small magnitudes. With the successful applications of the powerful ADMM optimization framework, existing research works \cite{zhang2018systematic, ren2019ADMMNN} achieve a very high weight reduction ratio while maintaining promising accuracy. However, non-structured methods lead to sparse and irregular weight matrices, which require indices to be stored in a compressed format.
Though saving the storage cost, the decoding of each stored index requires a search over the whole activation vector. Consequently, it suffers from limited acceleration in actual hardware implementation~\cite{han2017ese}.

\begin{figure}
    \vspace{-5mm}
    \centering
    \includegraphics[width=0.47 \textwidth]{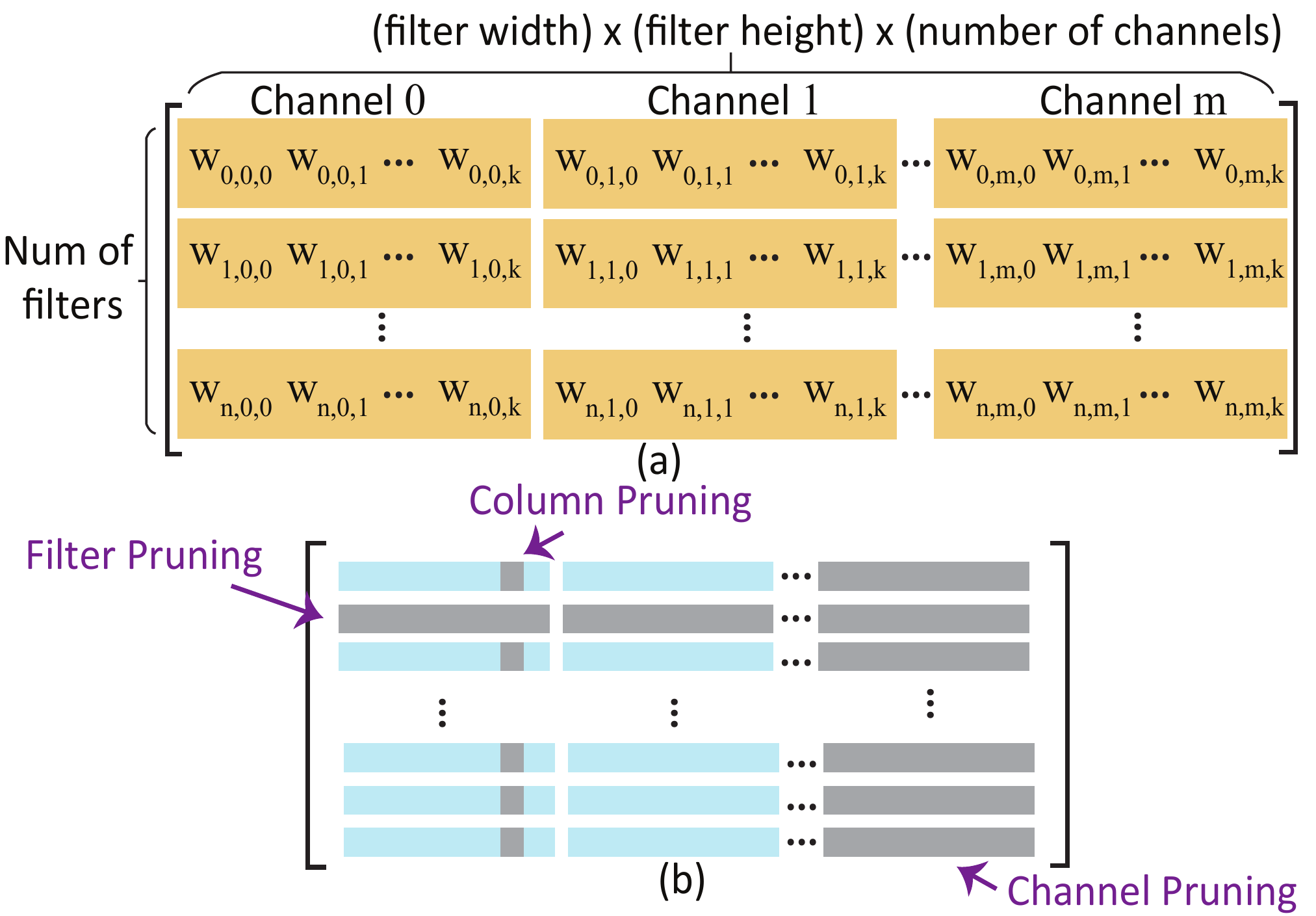}
    \caption{{\bf (a) To support GEMM computation, the weight tensor representation of a CONV layer is transformed into the weight matrix representation. (b) How different structured weight pruning schemes are implemented on the weight matrix representation.}}
    \label{fig:GEMM}
    \vspace{-7mm}
\end{figure}

\paragraph{Structured pruning} To overcome the limitations of non-structured pruning, recent works \cite{wen2016learning, he2017channel, min20182pfpce} considered to incorporate regularity in weight pruning with a main focus on convolutional (CONV) layers of DNNs. 
Previous works mainly focus on two types of structured pruning: filter pruning and channel pruning.
Filter pruning, also known as row pruning, removes the entire filter(s), while channel pruning removes the whole channel(s). 
Figure \ref{fig:GEMM} illustrates the example of transforming convolutional computation into general matrix multiplication (GEMM) by converting weight tensors and feature map tensors to matrices \cite{chetlur2014cudnn}.
In general, structured pruning directly reduces the dimension of a weight matrix and preserves a full matrix format, thereby facilitating hardware implementations. On the downside, the coarse-grained nature of structured pruning hurts the accuracy more significantly.

\subsection{DNN Acceleration on Mobile Devices}

Many efforts target accelerating DNN execution on mobile devices in the past few years, including 
MCDNN~\cite{han2016mcdnn},
DeepMon~\cite{huynh2017deepmon}, 
TFLite~\cite{TensorFlow-Lite}, TVM~\cite{chen2018tvm}, and Alibaba Mobile Neural Network~\cite{Ali-MNN}. 
However, most of them do not deeply exploit model compression techniques as RTMobile.
\textit{In particular, none of the existing frameworks can even support RNN acceleration on mobile devices.} 

\subsection{Research Motivation}
Based on the survey of recent research works, we conclude the following insights: (\romannumeral1) non-structured pruning has the advantage of very high compression ratio but is typically not compatible with GPU acceleration for inference; (\romannumeral2) structured pruning facilitates hardware implementations but is often subjected to accuracy degradation, especially when it is applied to time-based RNNs. 
To overcome the limitations of current methods, a more flexible and fine-grained pruning policy is needed. This work specifically focuses on RNN models that have not been extensively studied.




\section{Related Work}
\label{sec:related}

Many existing studies have implemented model compression algorithms for RNN acceleration on FPGAs \cite{li2015fpga,nurvitadhi2016accelerating,guan2017fpga, han2017ese,wang2018c,li2019rnn}. However, the majority of these works focus on constructing new RNN architectures \cite{nurvitadhi2016accelerating} rather than software and hardware co-design framework. Instead, our RTMobile proposes architecture designs in both software and hardware level. In this work, we mainly discuss and compare RTMobile with two most recent and related approaches, i.e., \textbf{ESE} \cite{han2017ese} and \textbf{C-LSTM} \cite{wang2018c}, which not only address the RNN model compression problem on algorithm/software but also take into account the hardware efficiency on hardware (i.e., FPGAs).

\subsection{ESE}
ESE proposes an optimized LSTM compression framework on FPGA, which sparses the model through parameter pruning \cite{han2015deep, han2015learning}. Compared with both CPU- and GPU-based implementations, ESE achieves higher energy efficiency on FPGA. However, the design of ESE has three main limitations: (1) ESE's irregular pruning method used for model compression causes large overhead when performing read/write operations on hardware; (2) the irregularity of weight matrix storage in ESE results in inefficient implementations of indices that consume extra storage cost, thus the computing power of the FPGA is not fully exerted; and (3) ESE only marginally improves compression ratio taking into account indices.

\subsection{C-LSTM}
In order to solve the problem caused by irregular pruning, Wang et al. \cite{wang2018c} propose an approach (called C-LSTM) to employ a structured compression technique using block-circulant matrices to compress the LSTM model. With regular structure of the block-circulant matrices, C-LSTM can further reduces both computational and storage complexity compared with ESE. However, the coarse-grained nature of structured pruning also cause relatively significant degradation on the model accuracy. Moreover, the advanced ADMM-based neural network pruning method, which can effectively handle both model compression and accuracy, is not supported in the C-LSTM training because it requires the most advanced optimizer in stochastic gradient decent (e.g., Adam optimizer).

\subsection{ADMM}
\label{sec:admm}
The pruning problem can be formulated as the minimization of $f(W,b) + g(W)$ by following: 
\begin{equation}
\small
\label{original}
\begin{aligned}
& \underset{ \{{\bf{W}}_{i}\}}{\text{minimize}}
& & f \big( \{\bf{W}_{i},{b}_{i}\}_{i=1}^N \big) + g \big( \{{\bf{W}}_{i}\}_{i=1}^N \big),
\\ & \text{subject to}
& & {\bf{W}}_{i}\in {\bf{\mathcal{S}}}_{i}, \; i = 1, \ldots, N,
\end{aligned}
\end{equation}
where N is the total number of weight tensor in recurrent neural network, $f(W,b)$ is the loss function,  and $g(W)$ is an indicator  function that is zero when the constraint S = \{ the number of nonzero weights is less than certain threshold \} is satisfied, but +$\infty$ otherwise.

The augmented Lagrangian formation of problem  (\ref{original}) is
\vspace{-0.50em}
\begin{equation}
\small
\begin{aligned}
\label{equ_lag}
 \textit{L}_p=\underset{ \{{\bf{W}}_{i}\} \}}{\text{minimize}}
\ \ \ & f \big( \{\bf{W}_{i},{b}_{i} \}_{i=1}^N \big) +  \sum_{i=1}^{N} \frac{\rho_{i}}{2}  \| {\bf{W}}_{i}-{\bf{Z}}_{i}+{\bf{U}}_{i} \|_{F}^{2}, \\
\end{aligned}
\end{equation}
where ${\rho_{i}}$ is a penalty value,  ${\bf{Z}}_{i}$ is pruning mask and ${\bf{U}}_{i}$ is dual variable.

The ADMM algorithm~\cite{boyd2011distributed} is to iteratively update the indicated pruning mask and retrain the neural network under this mask, until a good mask and neural network converge. It proceed by repeating iteration $k = 0, 1,\ldots$ as following: 
\vspace{-0.20em}
\begin{equation}
\small
    \bf{W}_{i}^{k+1} := \underset{ {\bf{W}}_{i}}{\text{arg min}}\quad \textit{L}_p(\{\bf{W}_i\}, \{\bf{Z}_i^k\}, \{\bf{U}_i^k\}),
\label{itera1}
\end{equation}
\vspace{-0.50em}
\begin{equation}
\small
    \bf{Z}_{i}^{k+1} := \underset{ {\bf{Z}}_{i}}{\text{arg min}}\ \textit{L}_p(\{\bf{W}_i^{k+1}\}, \{\bf{Z}_i\}, \{\bf{U}_i^k\}),
\label{itera2}
\end{equation}
\vspace{-0.50em}
\begin{equation}
\small
    \bf{U}_{i}^{k+1} := \bf{U}_{i}^{k}+\bf{W}_{i}^{k+1}-\bf{Z}_{i}^{k+1}.
\label{itera3}
\end{equation}
The pruning mask can be trained by Algorithm 1.



\begin{figure*}
    \vspace{-6mm}
    \centering
    \includegraphics[width=0.95 \textwidth]{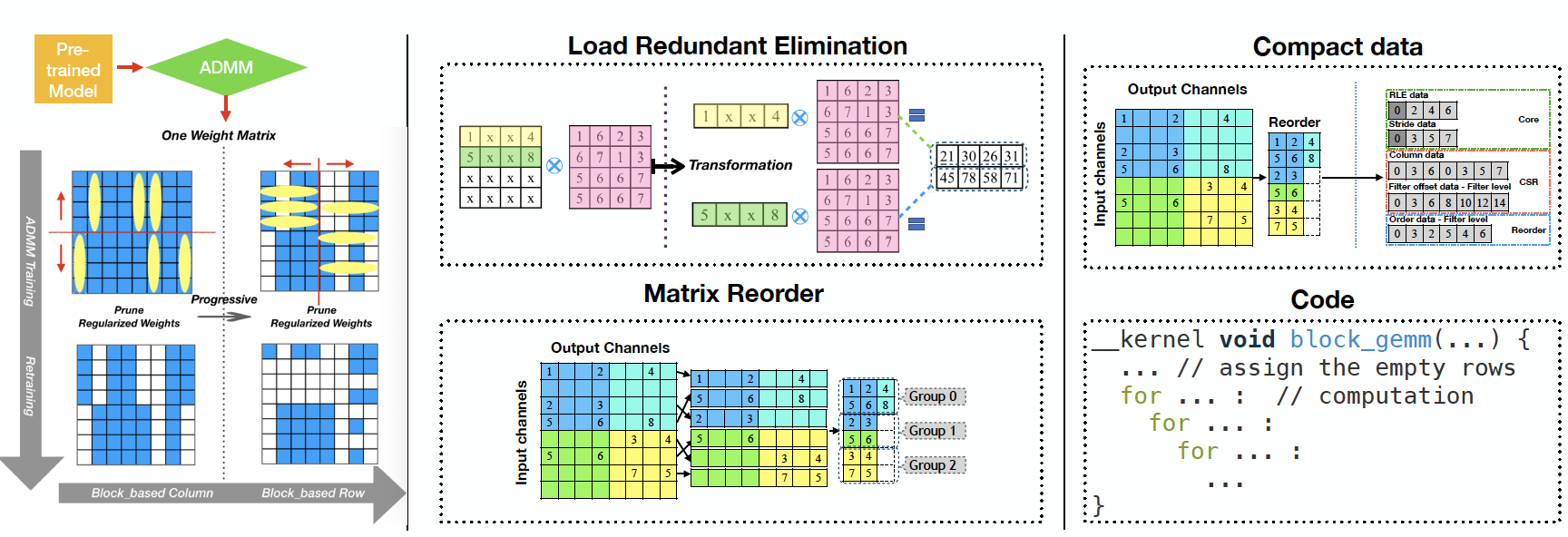}
    \vspace{-2mm}
    \caption{Systematic overview of RTMobile acceleration framework.}
    \label{fig:system_overview}
    \vspace{-2mm}
\end{figure*}

\section{Proposed RTMobile Framework}
\label{sec:rtmobile}


In this section, we describe in detail RTMobile, our proposed mobile acceleration framework for RNNs.

\subsection{Block-based Structured Pruning}
To better facilitate the compression ratio and ensure the structured model architecture for hardware implementations, we propose \textit{Block-based Structured Pruning (BSP)} algorithm. In general, training a BSP compressed model can be separated into two main steps: \textit{Step 1)} row-based column block pruning and \textit{Step 2)} column-based row pruning. 

The training process starts with splitting the whole weight matrix $\bm{W}$ into \textit{$Num_r$} rows horizontally. For each row, we divide it into \textit{$Num_c$} blocks and then perform the structured pruning using ADMM method (discussed in Section~\ref{sec:admm}). Then, we perform column-based row pruning over the entire weight matrix $\bm{W}$ in the step 2. Given the constraint of block number after dividing by \textit{$Num_c$} and \textit{$Num_r$}, the pruned model can achieve a satisfactory performance overhead on hardware.

The training process continues iteratively until all the blocks are pruned. We identify that by doing so, the training performance is stable, and the whole weight matrix after pruning is decentralized. Our BSP training approach is summarized in Algorithm 1.


\begin{figure}
\vspace{-2mm}
\begin{center}
	\includegraphics[width = 0.47\textwidth]{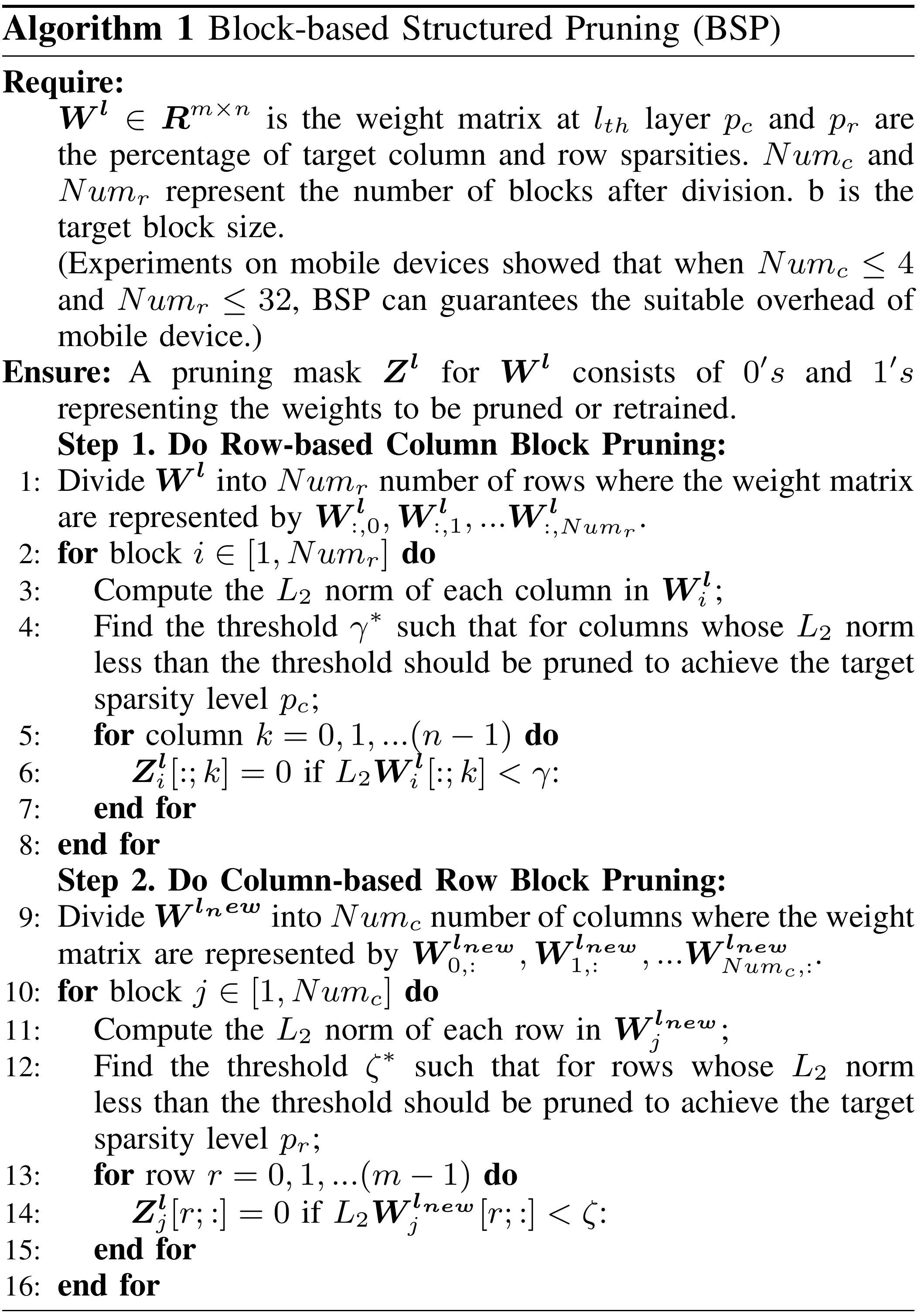}
	\vspace{-8mm}
\end{center}
\end{figure}

\subsection{Compiler-assisted RNN Acceleration Framework}

After block-based structured pruning, RTMobile relies on a compiler-assisted RNN acceleration framework to achieve efficient RNN inference on mobile devices. This compiler framework consists of three key optimizations that work on each RNN layer (as shown in Figure~\ref{fig:system_overview}): matrix  reorder, load redundancy elimination, and a compact data storage format for pruned RNN matrices, BSPC (i.e., Block-based Structured Pruning Compact format). These optimizations aim to address three key challenges in pruned RNN execution: {\em thread divergence} and {\em load imbalance} among threads, {\em redundant memory access}, and {\em unnecessary zero storage}. 


\paragraph{Matrix reorder}
The matrix is executed by multiple CPU/GPU threads simultaneously. Without a further reorder, these threads may execute rows with significantly divergent computations, causing severe load imbalance issue that hurts thread-level parallelism. Therefore, RTMobile introduces a matrix reorder optimization to group the rows with the same (or similar) computation patterns together. After this reorder, the rows in each group are assigned to multiple threads to achieve balanced processing.   

\paragraph{Redundant load elimination}
Within a group, each thread processes multiple continuous rows, offering us an opportunity of eliminating the redundant memory load operations. This optimization is specifically enabled by our block-based structured pruning, because after such pruning, the preserved weights in two neighbor rows may share the same pattern and require the same data in the input feature maps. It is difficult to explore this optimization opportunity for existing unstructured weight pruning due to its irregularity.   

\paragraph{BSPC format}
Our proposed block-based structured pruning also guides us to design a more compact data structure than traditional CSR format (called BSPC format) to store RNN weight matrices. This is because within each block the preserved weights only exist in certain rows and columns, enabling to further compact the index array in CSR. The BSPC format also includes the matrix reorder information to match the corresponding input feature map with the weight matrix. The BSPC format significantly reduces the memory footprint thus alleviating the memory-bound issue in RNN execution.

In addition to above optimizations, our compiler framework also includes an auto-tuning component to perform an offline search of the best execution configurations like the matrix tiling size, unrolling size, memory placement, etc. In particular, we employ it to find the best block size that results in an optimal combination of accuracy and performance. 

\begin{table*}[tbp]
\small
\caption{\textbf{Results of Different Model Compression Methods on GRU Using TIMIT Dataset:} PER is {\em phone error rate}, the lower the better. Baseline PER is for dense (non-pruned) models and pruned PER is for pruned compressed models. PER Degrad. represents for the PER degradation, i.e., $PER_{pruned} - PER_{baseline}$. The rest columns show the column compression rate, row compression rate, the number of preserved parameters, and the overall compression rate, respectively.}
\vspace{-2mm}
\begin{center}
\label{Table: Prune_software}
\scalebox{0.8}{
\begin{tabular}{|l|c|c|c|c|c|c|c}
\hline
Method  & \begin{tabular}[c]{@{}c@{}}PER (\%)\\ (baseline - pruned) \end{tabular} & PER Degrad. & Column Compress. Rate & Row Compress. Rate & Para. No. & Overall Compress. Rate \\
\hline\hline

ESE\cite{han2017ese} & 20.40 - 20.70 & 0.30 & -- & -- & 0.37M & 8$\times$ \\\hline
C-LSTM\cite{wang2018c} & 24.15 - 24.57 & 0.42 & -- & -- & 0.41M & 8$\times$ \\\hline
C-LSTM\cite{wang2018c} & 24.15 - 25.48 & 1.33 & -- & -- & 0.20M & 16$\times$ \\\hline
BBS\cite{cao2019efficient} & 23.50 - 23.75 & 0.25 & -- & -- & 0.41M & 8$\times$ \\\hline
Wang\cite{wang2019acceleration} & -- & 0.91 & -- & -- & 0.81M & 4$\times$ \\\hline

E-RNN\cite{li2019ERNN} & 20.02 - 20.20 & 0.18 & -- & -- & 1.20M & 8$\times$ \\\hline\hline
\textbf{BSP (ours)}  & 18.80 (w/o pruning)  &  0  &  1  & 1  &  9.6M & 1$\times$ \\\hline
\textbf{BSP (ours)}  & 18.80 - 18.80 &  0  &  10  & 1  &  0.96M & 10$\times$ \\\hline
\textbf{BSP (ours)} & 18.80 - 19.40 &  0.60  &  16 &  1.25  &  0.48M & 19$\times$ \\\hline
\textbf{BSP (ours)} & 18.80 - 19.60 &  0.80  &  16 &  2  &  0.33M & 29$\times$ \\\hline
\textbf{BSP (ours)} & 18.80 - 20.60 &  1.80  &  16 &  5  &  0.22M & 43$\times$ \\\hline
\textbf{BSP (ours)} & 18.80 - 21.50 &  2.70  &  20 &  8  &  0.12M & 80$\times$ \\\hline
\textbf{BSP (ours)} & 18.80 - 23.20 &  4.40  &  16 &  16  &  0.09M & \textbf{103$\times$} \\\hline
\textbf{BSP (ours)} & 18.80 - 24.20 &  5.40  &  20 &  10  &  0.06M & 153$\times$ \\\hline
\textbf{BSP (ours)} & 18.80 - 24.20 &  5.40  &  20 &  16  &  0.04M & \textbf{245$\times$} \\\hline
\textbf{BSP (ours)} & 18.80 - 25.50 &  6.70  &  20 &  20  &  0.03M & \textbf{301$\times$} \\\hline
\end{tabular}}
\end{center}
\vspace{-2mm}
\end{table*}

\begin{table*}[tbp]
\small
\caption{\textbf{Performance and Energy Evaluation on Mobile GPU and CPU:} GOP refers to Giga Operations. GPU/CPU energy efficiency is calculated as $Inference Frames/(Power\times Inference Time)$, i.e., the number of frames inferred per unit energy consumption. This table normalizes our method's GPU/CPU energy efficiency by the ESE FPGA implementation's. As our compression rate reaches $245\times$, our GPU inference time becomes slightly faster than ESE's (82.7us). Our GPU implementation uses 16-bit floating point.}
\vspace{-2mm}
\begin{center}
\label{Table: Accer_hardware}
\scalebox{0.8}{
\begin{tabular}{|l|c|c|c|c|c|c|c|c}
\hline
Compression Rate  &  GOP & GPU Time / Frame (us) & GPU GOP/s & \begin{tabular}[c]{@{}c@{}}GPU Energy Efficiency\\ (normalized with ESE)\end{tabular} & CPU Time / Frame (us) & CPU GOP/s & \begin{tabular}[c]{@{}c@{}}CPU Energy Efficiency\\ (normalized with ESE)\end{tabular}\\
\hline\hline

1$\times$ (baseline)  & 0.5800 & 3590.12 & 161.55 & 0.88 & 7130.00 & 81.35 & 0.25 \\\hline
10$\times$ & 0.0580 & 495.26 & 117.11 & 6.35 & 1210.20 & 47.93 & 1.48 \\\hline
19$\times$ & 0.0330& 304.11 & 108.51 & 10.35 & 709.33 & 46.52 & 2.52 \\\hline
29$\times$ & 0.0207 & 233.89 & 88.29 & 13.45 & 464.73 & 44.43 & 3.85 \\\hline
43$\times$ & 0.0143 & 186.05 & 76.86 & 16.91 & 344.77 & 41.48 & 5.19 \\\hline
80$\times$& 0.0080 & 130.00 & 61.54 & 24.2 & 218.01 & 36.70 & 8.20 \\\hline
103$\times$ & 0.0060 & 109.76 & 54.66 & 28.67 & 202.72 & 29.59 & 8.82 \\\hline
153$\times$ & 0.0039 & 97.11 & 40.16 & 32.4 & 170.74 & 22.84 & 10.47 \\\hline
\textbf{245$\times$} & \textbf{0.0028} & \textbf{81.64} & \textbf{34.30} & \textbf{38.54} & \textbf{151.28} & \textbf{18.51} & \textbf{11.82} \\\hline
301$\times$ & 0.0020 & 79.13 & 25.27 & 39.76 & 145.93 & 13.71 & 12.25 \\\hline
\end{tabular}}
\end{center}
\vspace{-4mm}
\end{table*}

\begin{figure}[t]
\begin{center}
	\includegraphics[width = 0.4\textwidth]{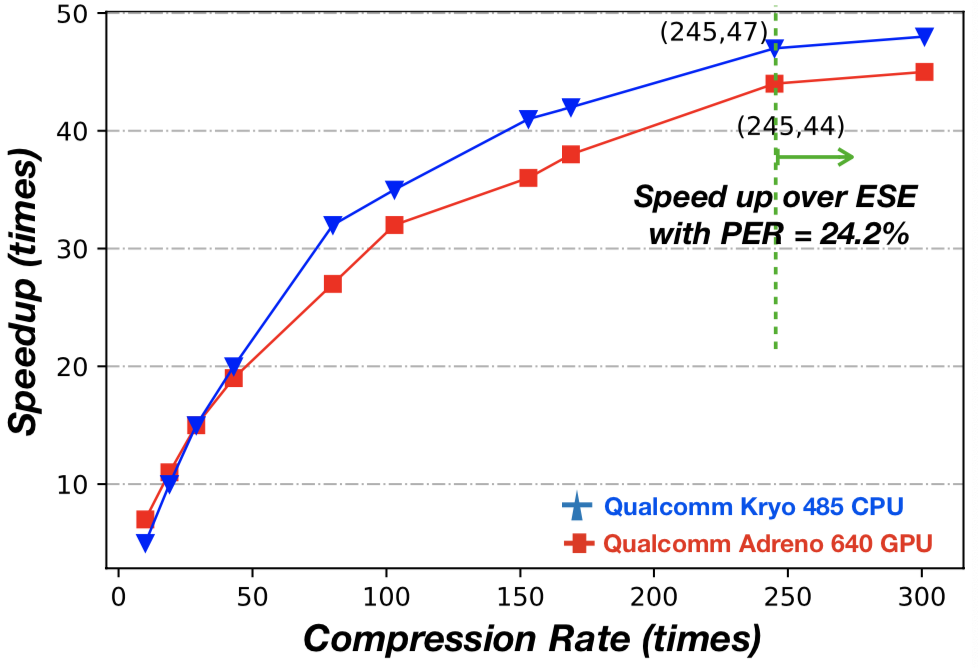}
	\caption{Speedup using RTMobile with different compression rates on mobile platform.}
	\label{fig:mobile_acc}
\end{center}
\vspace{-3mm}
\end{figure}

\section{Experimental Evaluation}
\label{sec:evaluation}

In this section, we evaluate RTMobile by comparing it with several state-of-the-art methods. There are three evaluation objectives: 1) comparing RTMobile with other model compression methods and demonstrating that our method outperforms others in both compression rate and accuracy; 2) showing RTMobile has both higher computational efficiency and energy efficiency than a well-known deployment on FPGA (ESE\cite{han2017ese})\footnote{We compare RTMobile on mobile with ESE on FPGA because (1) \textit{none of the existing RNN acceleration works supports mobile device}, and (2) ESE provides one of the highest inference accuracy among prior works.}; and 3) studying the relationship between compression rate and inference execution time.

\subsection{Experiment Setup}
\noindent\textbf{Experimental Platform.}
We conduct our experiments using a Samsung Galaxy S10 with the latest Qualcomm Snapdragon 855 mobile platform, which consists of a Qualcomm Kryo 485 Octa-core CPU and a Qualcomm Adreno 640 GPU.

\noindent\textbf{Model Architecture.}
We evaluate RTMobile and compare it with the state-of-the-art methods on the popular GRU RNN model, which has been widely used in previous studies \cite{han2017ese, wang2018c, li2019ERNN}. 
The GRU model contains 2 GRU layers and about 9.6M overall number of parameters.

\noindent\textbf{Evaluation Dataset.}
We conduct our experiments on the TIMIT dataset \cite{garofolo1993timit}, which is widely adopted for evaluating automatic speech recognition systems. The TIMIT dataset contains broadband recordings from 630 speakers reading ten phonetically rich sentences in eight major dialects of American English, each reading ten phonetically rich sentences. 

\subsection{Evaluation Results and Discussion}


\noindent{\bf Compression Rate and Accuracy.} Table \ref{Table: Prune_software} illustrates the results (including phone error rate and number of preserved parameters) of RTMobile with different compression rates and the comparison with other state-of-the-art methods, including ESE\cite{han2017ese}, C-LSTM\cite{wang2018c}, BBS\cite{cao2019efficient}, Wang\cite{wang2019acceleration} and E-RNN\cite{li2019ERNN}. For a fair comparison, we train all models using the same TIMIT dataset\cite{garofolo1993timit}. 
Benefit from the most advanced \textit{PyTorch-Kaldi Speech Recognition Toolkit}\cite{pytorch-kaldi}, the baseline GRU model for our RTMobile can achieve higher recognition accuracy than the other methods before pruning, e.g., our PER is 5.35\% lower than C-LSTM's (18.80\% v.s. 24.15\%).

We observe that our proposed BSP method can guarantee no accuracy degradation when the compression rate is not higher than $10\times$, which is superior than ESE and C-LSTM from both compression rate and inference accuracy. 
We also observe that BSP can stably keep a high accuracy compared to the other methods when the compression rate is relatively high. For instance, when the compression rate is \textbf{103$\times$}, the BSP pruned model can even outperform the C-LSTM baseline model in terms of both compression rate and accuracy. 
The C-LSTM baseline model (with 3.25M parameters) has $36\times$ more parameters than our BSP pruned model, but its PER is 0.95\% higher than ours (24.15\% vs. 23.20\%).
In addition, we use BSP to further prune the model until the rate of \textbf{301$\times$} and observe that our method can well adapt to \textit{ultra-high compression rate} scenario. 
For example, our model with \textbf{245$\times$} compression rate can still maintain the same-level PER \textbf{}as the C-LSTM baseline model (24.20\% vs. 24.15\%) and reduce the  parameter number by over $80\times$ (0.04M vs. 3.25M).

\noindent{\bf Inference Time and Energy Efficiency.} Table \ref{Table: Accer_hardware} presents the evaluation results of RTMobile's inference time, Giga Operations Per Second (GOP/s), and energy efficiency (normalized with ESE method) on mobile GPU and CPU, respectively. 
The table illustrates that, when the compression rate is higher than \textbf{245$\times$}, RTMobile can outperform in energy efficiency by about $40\times$ compared with ESE while maintaining the same inference time (ESE's inference time is 82.7 us) on the mobile GPU (ESE uses a large FPGA platform of 41W power, and thus it is easier to achieve higher energy efficiency than speed). Please note that this is a clear feat, as it is typically perceived that FPGA is more energy-efficient than general-purpose computing devices. This is because of two main reasons. First, comparing to ESE's activation calculation by look-up tables that results in limited parallelization and irregular memory accesses (two key performance factors on FPGA), RTMobile's compiler optimizations significantly improve both the parallelization and memory performance. Second, RTMobile has a much better compression rate (with a negligible accuracy loss), resulting in a more significant computation reduction. Although our compression rates are significant, we must emphasize that the inefficiency in FPGA implementation in ESE (especially activation) plays an equally important, if not more, role. As can be seen from the table, our GPU energy efficiency (frames in unit energy) is almost the same as ESE (which uses compression) even when we do not have any pruning. 
With increase in the compression rate, the computation pattern becomes I/O and memory bounded, the memory access pattern becomes more irregular, which leads to lower CPU/GPU GOP/s.

\noindent{\bf Relationship between Compression Rate and Inference Time.} Figure \ref{fig:mobile_acc} further illustrates the relationship between inference time and compression rate. The inference time is in the form of speedups over our own dense CPU/GPU baselines, respectively. The speedup grows as compression rate increases. The speedup becomes stable when compression rate reaches to a certain range (e.g., compression rate reaches \textbf{250$\times$}). When the compression rate is \textbf{245$\times$}, our inference time on mobile GPU is the same to ESE's on FPGA.

\section{Conclusion}
\label{sec:conclusion}

In this paper, we propose \textit{the first RNN acceleration framework for mobiles}, called RTMobile. We develop a novel block-based pruning algorithm and three compiler optimizations to achieve real-time inference without any accuracy degradation.  
Experimental results demonstrate that RTMobile significantly outperforms the existing RNN hardware acceleration methods in terms of compression rate, inference accuracy, execution time, and energy efficiency. 

\bibliographystyle{IEEEtran}                            
\bibliography{reference}

\end{document}